\documentclass[prb,citeautoscript,showpacs,twocolumn,amsmath,amssymb,preprintnumbers,floatfix]{revtex4}

\usepackage{amsmath,amssymb}
\usepackage{bm}
\usepackage{graphicx}
\usepackage{ae}
\usepackage{epsfig}
\usepackage{dcolumn}

\newcommand{\bfk}{ {\bf k}} 

\newcommand{\ki}[1]{\bfk_{#1}}

\newcommand{\bracket}[3]{\ensuremath{\langle #1 | #2 | #3 \rangle}}

\begin{document}

\title{{\rm\small\hfill (accepted at Appl.\ Phys.\ Lett.)}\\
        Auger recombination rates in nitrides from first principles}

\author{Kris T. Delaney}
\affiliation{Materials Research Laboratory, University of California, Santa Barbara, CA 93106-5121}

\author{Patrick Rinke}
\affiliation{Materials Department, University of California, Santa Barbara, CA 93106-5050}

\author{Chris G. Van de Walle}
\affiliation{Materials Department, University of California, Santa Barbara, CA 93106-5050}

\date{\today}

\begin{abstract}
We report Auger recombination rates for wurtzite InGaN calculated from first principles density-functional and many-body-perturbation theory. Two different mechanisms are examined -- inter- and intra-band recombination -- that affect different parts of the emission spectrum. In the blue to green spectral region and at room temperature the Auger coefficient can be as large as 2$\times$10$^{-30}$cm$^{6}$s$^{-1}$; in the infrared even larger. 
Since Auger recombination scales with the cubic power of the free-carrier concentration it becomes an important non-radiative loss mechanism at high current densities. Our results indicate that Auger recombination may be responsible for the loss of quantum efficiency that affects InGaN-based light emitters.
\end{abstract}


%


\maketitle


Indium gallium nitride (InGaN) alloys are now already being used for 
light emitting and laser diodes in the green to ultraviolet part of the  
spectrum\cite{NakamuraBook:2000}, but increases in 
internal quantum efficiency (IQE) are still required to allow broader applications. 
The IQE of InGaN devices is limited by loss mechanisms that, at high  
drive currents (i.e., high carrier concentrations) lead to a decrease in IQE, a phenomenon
commonly referred to as ``efficiency droop''.
The precise nature of these loss mechanisms has been the subject of intense debate, 
and a variety of candidates have been proposed (see Ref.\; \onlinecite{Shen/etal:2007}).
Recently, Shen {\it et al.} suggested Auger recombination as the dominant 
source \cite{Shen/etal:2007,Gardner/etal:2007}.
Loss due to Auger recombination scales with the cubic power of the free-carrier density 
and would thus dominate at the high carrier concentrations at which the reduction in IQE is 
observed. While Shen {\it et al.} found a cubic dependence of the IQE on the free-carrier 
concentration in optically pumped InGaN LED devices\cite{Shen/etal:2007}, it is difficult to
discriminate between different radiationless processes experimentally. 
In this Letter we demonstrate by means of rigourous first-principles calculations, 
in which the Auger process can explicitly be isolated, that
Auger recombination is indeed an important loss mechanism in wurtzite InGaN.

\begin{figure}
  \begin{center} 
    \epsfig{bbllx=124,bblly=147,bburx=626,bbury=514,clip=,file=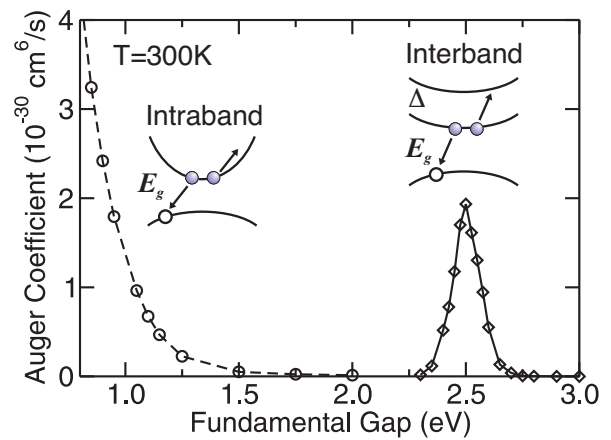,width=0.90\columnwidth}
    \caption{\label{fig:Aug_c} \emph{eeh} Auger coefficient for light holes
	  as a function of fundamental band gap 
	  for a simulated InGaN 
      alloy (see text) at $T=300$\,K and $n=1\times10^{19}$\,cm$^{-3}$. 
        The statistical error bars of the Monte Carlo integration are smaller
        than the symbols for all data points presented.
      Intraband Auger, which involves scattering of an electron to an unoccupied part
      of the lowest conduction band (left inset, dashed line), dominates in In-rich alloys. However, for alloy
      concentrations relevant for solid-state lighting, interband Auger recombination, which
      involves scattering to the second-lowest conduction band (right inset, solid line), is dominant. 
              }
  \end{center}
\end{figure}

In the direct Auger process, an electron recombines with a hole, 
but instead of emitting a photon the process results in the excitation of another carrier to
a higher-energy state (see insets of Fig.~\ref{fig:Aug_c}).
This can also be viewed as two electrons colliding in the vicinity of a hole, 
resulting in a radiationless \emph{e-h} recombination event, the energy and momentum 
of which is absorbed by the second electron
(\emph{eeh} process). Alternatively, the \emph{hhe} process involves two holes and one electron.
A first-principles description of the Auger recombination rate
therefore requires an accurate calculation of the band structure and the transition 
probabilities for all relevant \emph{eeh} and \emph{hhe} processes as inputs.

\begin{figure*}
  \begin{center} 
    \epsfig{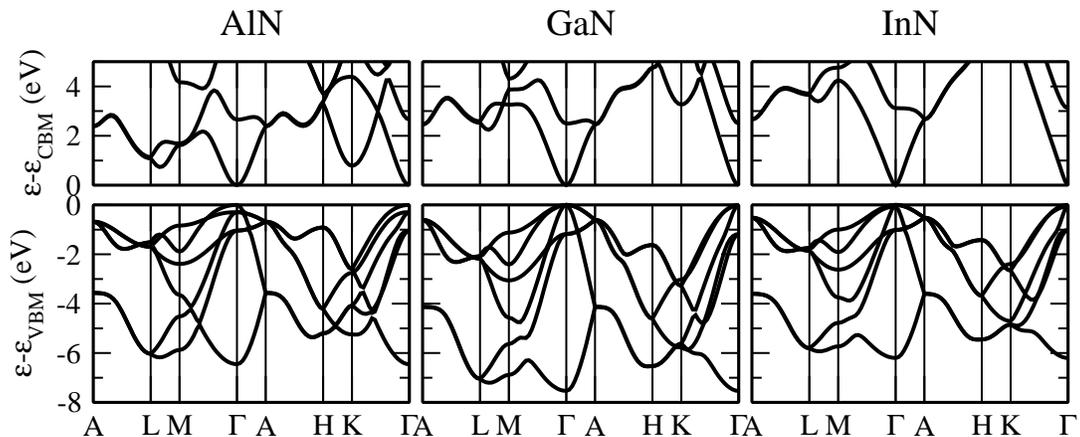}
    \caption{\label{fig:BS} OEPx(cLDA)+$G_0W_0$ band structure for wurtzite AlN, GaN and InN. 
              For ease of comparison the
              valence bands have been aligned at the valence-band maximum (VBM) and the
	      conduction bands at the conduction-band minimum (CBM).
              }
  \end{center}
\end{figure*}

We obtain the band structures of the host materials  
by combining density-functional theory (DFT) with
many-body perturbation theory in the $G_0W_0$ approximation\cite{Hedin65}. 
This approach accurately describes band structures of solids as measured
by direct and inverse 
photoemission\cite{Aulbur/Jonsson/Wilkins:2000,Onida/Reining/Rubio:2002,Rinke/etal:2005,Rinke/pssb}.  Combined with 
the exact-exchange optimized effective potential approach (OEPx(cLDA)) for the DFT
part, it produces accurate band structures for AlN, GaN and InN in their zinc-blende and 
wurtzite phases\cite{Rinke/etal:2006,Rinke/etal:2008}. 
We do not include spin-orbit coupling, which has only very small effects on the band 
structure of nitrides. The results for wurtzite are
shown in Fig.~\ref{fig:BS}.

The Auger rate $R$ is given by\cite{Laks/etal:1990,Haug/etal:1978} 
\begin{alignat}{1}
  R=2\frac{2\pi}{\hbar}\frac{V^3}{(2\pi)^9}  \iint & \negthickspace \negmedspace \iint
  \left|M_{1,2,3,4}\right|^2 P_{1,2,3,4}   \nonumber \\
 &  \delta(\ki{\rm sum})  
   \delta(E_{\rm sum}) \: d\ki{1}d\ki{2}d\ki{3}d\ki{4}
\label{eq:R}
\end{alignat}
where states $3$ and $4$ are, e.g. for \emph{eeh},
electrons in the conduction band and 
states $1$ and $2$ are a hole in the valence band and an
electron in a higher-energy conduction-band state. $\ki{\rm sum}$ and $E_{\rm sum}$ are short for $\ki{1}+\ki{2}-\ki{3}-\ki{4}$ and $E_1+E_2-E_3-E_4$, respectively.
The statistics factor
\begin{equation}
\label{eq:P}
P_{1,2,3,4}=\left(1-f\left(E_1\right)\right)\left(1-f\left(E_2\right)\right)
                     f\left(E_3\right)f\left(E_4\right)
\end{equation}
is determined by the Fermi occupation functions $f$, and $M_{1,2,3,4}$
are the Auger matrix elements of the screened Coulomb potential $W$ 
\begin{equation}
\label{eq:M}
  M_{1,2,3,4}=\bracket{\phi_1\phi_2}{W}{\phi_3\phi_4} + \mathrm{EX},
\end{equation}
where $\mathrm{EX}$ denotes exchange terms, as detailed in Ref.\; \onlinecite{Ridley:1982}.

We compute $M_{1,2,3,4}$ directly with the OEPx(cLDA) wave functions 
and the non-local, dynamically screened Coulomb interaction $W_0$ from an RPA calculation. 
Our OEPx(cLDA) calculations were performed with the \texttt{S/PHI/nX} plane-wave
pseudopotential code\cite{SPHInX}, while for the $G_0W_0$ 
calculations we have employed the \texttt{gwst} space-time 
code\cite{GW_space-time_method:1998,GW_space-time_method_enh:2000,GW_space-time_method_surf:2007}. 
With all operators being treated in real space, we have adapted the \texttt{gwst} code  
to compute the matrix elements of
the screened Coulomb interaction in Eq.~\ref{eq:M}.
In contrast with earlier work for semiconductors, the Auger matrix elements have been 
determined entirely from first principles.
The calculations were performed at the experimental lattice constants reported 
in Ref.\; \onlinecite{Rinke/etal:2008}.
For additional technical details and convergence
parameters we refer to Refs.\; \onlinecite{Rinke/etal:2005} and\; \onlinecite{Rinke/etal:2006}. We also note that
a plane-wave cutoff of $17$\,Ha is required for the matrix elements such that the Auger rate is converged to within
10\%.

For both $eeh$ and $hhe$ Auger processes we distinguish between \emph{intra}- and 
\emph{inter}-band events, depending on whether the final state of the scattered carrier 
lies in the same band as the initial state (see insets of Fig.~\ref{fig:Aug_c}).
For typical concentrations of injected carriers (10$^{17}$--10$^{20}$ cm$^{-3}$) holes and 
electrons are confined to a small region of the Brillouin zone around the $\Gamma$ point. 
Assuming equal electron and hole concentrations 
the maximum momentum transfer to the Auger electron or hole that is
scattered into the final state is $3k_F$ (at $T=0$\,K), where $k_F$ is the Fermi momentum). 
Since energy and momentum have to be conserved in the Auger process,
intraband Auger events are therefore only likely to occur for alloy 
compositions with small band gaps, as we demonstrate later. 
For larger band gaps intraband processes are negligible, and the possibility of
interband Auger processes then depends on whether other valence or conduction bands are
available into which holes or electrons can be scattered while energy is conserved.  Our accurate
band structures allow us to determine that possibility.

Focusing first on $eeh$ processes, we note that
the lowest conduction band in all three binary wurtzite phases is nondegenerate and
almost spherically symmetric around the $\Gamma$-point. The next higher conduction 
band is situated at an energy $\Delta$ above the CBM at $\Gamma$ ranging from 2.5 to 3.1~eV  
(see Fig.~\ref{fig:BS} and Table~\ref{tab:lat_gap}).  This strongly suggests that interband 
$eeh$ Auger processes that excite an electron into this 
second conduction may occur for InGaN alloys with matching band gaps.  
This second conduction band was not included in a recent $k.p$ study 
by Hader {\it et al.}, leading them to 
conclude that direct Auger losses are negligible in InGaN quantum wells \cite{Hader/etal:2008}.

The formidable computational challenges involved in calculating Auger rates 
were discussed by Laks {\it et al.}\cite{Laks/etal:1990}
The calculation of the Auger recombination rate itself [Eq.~(\ref{eq:R})] 
scales with the fourth power of the number of $k$ points, rendering explicit integration over
a grid of $k$ points prohibitively expensive.  We therefore tackled this multidimensional
integration with a Monte Carlo approach. We compute statistical averages over 40,000,000
Monte Carlo steps, chosen so that the error of the mean is always at least one order of magnitude 
lower than the value of the mean. 
Because of the expense involved in a first-principles calculation of Auger matrix elements
[Eq.~(\ref{eq:M})], we confined the calculations
to a finite number of points in the twelve-dimensional $k$-space (10,000 elements in total), 
and employed a linear-interpolation scheme for $k$ points off the mesh. 
This approach is justified because we found that the matrix
elements vary only weakly over the small volume of the Brillouin zone involved in Auger recombination.
Accurate interpolations of the OEPx(cLDA)+$G_0W_0$ band structures 
were obtained using an anisotropic effective-mass model\cite{Rinke/etal:2008}. 

Our present calculations are aimed at examining Auger rates for a wide range of
InGaN band gaps.  Explicit evaluation of band structures and wave functions for each of the
corresponding alloy compositions would be prohibitive.  We
performed calculations entirely from first-principles for $eeh$ processes in 
pure GaN, and modeled different
alloy compositions by applying a ``scissor shift''
to the band gap (i.e., rigidly shifting all conduction bands relative to the valence bands).  
While this choice is not optimal for calculations of alloys with large In concentrations, 
we note that In concentrations in current optoelectronic devices do not exceed 15\% and thus an extrapolation of band
parameters and matrix elements from GaN is the most sensible choice. 
With each $E_g$ value we associate an alloy composition by using a bowing parameter 
$b=2.5$\,eV\cite{Stepanov/etal:2001}, and then use linear interpolation 
to obtain $\Delta$ based on the values listed in Table~\ref{tab:lat_gap}.
This approach is justified by the similarity of the wave functions
and band structure for corresponding bands of GaN and InN. 

\begin{table}
  \begin{ruledtabular}
    \begin{tabular}{lddd|ddd}
     & \multicolumn{3}{c|}{wurtzite} & 
          \multicolumn{3}{c}{zinc blende}\\
     & \multicolumn{1}{c}{AlN} & 
	  \multicolumn{1}{c}{ GaN} & 
          \multicolumn{1}{c|}{InN} & 
          \multicolumn{1}{c}{AlN} & 
          \multicolumn{1}{c}{GaN}  &
          \multicolumn{1}{c}{InN} \\
      \hline
      $E_{g}$                      &  6.47  &  3.24  &  0.69  &   6.53 &  3.07 &  0.53  \\
      $\Delta$                       &  2.65  &  2.50  &  3.12  &  8.62 & 9.07 &	10.12	     \\ 
      $\Delta_c$              & -0.295 &  0.034 &  0.066 &
    \end{tabular} 
  \end{ruledtabular}
  \caption{\label{tab:lat_gap}\small Band gap ($E_{g}$), gap between the first and the second
           conduction band at $\Gamma$ ($\Delta$) and crystal-field splitting ($\Delta_c$) 
             calculated with the OEPx(cLDA)+$G_0W_0$ approach for AlN, GaN and InN.}
\end{table}

The Auger coefficient, defined as $C=R/n^3$, is reported in Fig.~\ref{fig:Aug_c} as a
function of the fundamental band gap for $T=300$\,K and a carrier density 
of $1\times10^{19}$\,cm$^{-3}$. 
For the purposes of the figure, we populated only the light-hole band with holes. 
Holes in other valence bands lead to a similar magnitude of the Auger coefficient.
In our calculations, which use Fermi-Dirac statistics,
the Auger coefficient is largely independent of $n$ for densities up to $1\times10^{19}$\,cm$^{-3}$.
We note that the interband process has a weak dependence on the choice of bowing parameter due to 
the method of choosing $\Delta$. For $b$ ranging from $1$ to $4$\,eV, the center of the recombination peak
varies by less than $0.1$\,eV.

Three distinctly different regimes emerge. For large band gaps, 
intraband Auger recombination is negligible and
interband recombination dominates. For small band gaps ($< 1.0$ eV) the situation
is reversed. 
In an intermediate energy region, both \emph{eeh} recombination processes
are irrelevant. The two Auger regimes exhibit a different characteristic behavior as a function
of band gap:
while interband recombination peaks when $\Delta$ and $E_{g}$ are in
resonance, intraband recombination monotonically increases with decreasing band gap.
$hhe$ processes are not included here, but inspection of the band structures in Fig.~\ref{fig:BS}
allows us to conclude that they do not contribute at $E_g$ values above 2.5 eV.

The Auger coefficients computed with our model fall within the (very wide) 
experimentally reported range of 
1$\times$10$^{-34}$ to 5$\times$10$^{-28}$ cm$^{6}$s$^{-1}$ (Ref.\; \onlinecite{Shen/etal:2007}).  
In particular, they agree well with the values reported by
Shen {\it et al.}\cite{Shen/etal:2007} based on their optical pumping experiments.
Their reported Auger coefficient ranged from (1.4 - 2.0)$\times$10$^{-30}$ cm$^{6}$s$^{-1}$ and
increased with In concentrations rising from 9 to 15\%.
Our calculated values in Fig.~\ref{fig:Aug_c} are in qualitative and quantitative agreement with 
these observations, strengthening the case that the observed losses are indeed
due to Auger recombination.
Interestingly, Fig.~\ref{fig:Aug_c} shows that the Auger coefficient continues to increase 
when the band gap is lowered down to ~2.5 eV; i.e., for longer wavelengths.  This indicates that
rising Auger losses could well play an important role in the ``green gap'' problem, the
well known decrease in efficiency of InGaN light emitters at longer emission wavelengths.
  
Based on our results we can examine possible strategies for reducing Auger losses in the blue/green 
region of the spectrum.  
(1) Using the zinc-blende phase, in which the second conduction band 
occurs at much higher energies (see Table~\ref{tab:lat_gap}); however, growth of 
high quality phase-pure zinc-blende nitrides has proven very difficult. 
(2) Strain engineering the band structure
to move the second conduction band away from the resonance; however, both the first and second 
conduction band are nondegenerate and thus sensitive only to hydrostatic strain, and our calculated
deformation potential for the energy difference $\Delta$ is quite small.
(3) Tuning InGaAlN alloy compositions.  Again, this is likely to be fruitless
because the value of $\Delta$ is similar for all three nitrides (Table~\ref{tab:lat_gap}). 

In summary, we have presented first-principles evaluations of Auger recombination rates 
in InGaN alloys, showing the presence of a resonance that leads to increasing values of the
Auger coefficient for wavelengths ranging from blue to green.  The calculated values are in good
agreement with experiment and confirm that Auger recombination is a key loss mechanism in nitride
light emitters.

This work was supported by DOE and 
Inlustra Technologies, LLC under Award No.~DE-FC26-07NT43228 (Program Manager: B. Dotson).
It made use of MRL Central Facilities supported by the NSF MRSEC Program under award No.~DMR05-20415. P. Rinke acknowledges the support of the Deutsche Forschungsgemeinschaft.
We acknowledge fruitful discussions with C. Weisbuch, P. Yu, Yan Qimin, 
Yifan Huang and A. Janotti.  

\newpage


\end{document}